\documentclass[11pt]{article}
\usepackage{fullpage}
\usepackage{times}
\usepackage{boxedminipage,url}
\usepackage{wrapfig}

\usepackage{amsmath,amssymb,amsfonts,amsthm} 
\usepackage{graphicx}
\usepackage{subfigure}

\newcommand{\bvec}[1]{{\bf #1}}
\newcommand{\junk}[1]{}

\newcommand{\todo}[1]{{\bf ***TODO: #1***}}

\newtheorem{theorem}{Theorem}
\newtheorem{lemma}{Lemma}

\newtheorem{proposition}{Proposition}

\newtheorem{definition}{Definition}

\newcommand{\X}{\Sigma}
\newcommand{\Xn}{\Sigma^n}
\newcommand{\state}{q}
\newcommand{\states}{Q}

\newcommand{\fil}{\Phi}
\newcommand{\polylog}{\text{polylog}}
\newcommand{\seq}{\bvec{x}}

\newcommand{\cleanup}{\eta}

\newcommand{\mud}{\text{MUD}}
\newcommand{\amud}[1]{#1\text{-MUD}}
\newcommand{\pmud}{\text{pMUD}}
\newcommand{\rmud}{\text{rMUD}}
\newcommand{\imud}{\text{iMUD}}

\newcommand{\stream}{\text{SS}}
\newcommand{\astream}[1]{#1\text{-SS}}
\newcommand{\rstream}{\text{rSS}}
\newcommand{\pstream}{\text{pSS}}
\newcommand{\istream}{\text{iSS}}

\newcommand{\tree}{{\cal T}}

\newcommand{\one}{{\tt a}}
\newcommand{\two}{{\tt b}}

\newcommand{\set}[1]{ \{ {#1} \} }


\title{On the Complexity of Processing\\ Massive, Unordered, Distributed Data}

\author{Jon Feldman\thanks{Google, Inc., New York, NY. {\tt jonfeld@google.com}.}
\and S. Muthukrishnan\thanks{Google, Inc., New York, NY. {\tt muthu@google.com}.}
\and Anastasios Sidiropoulos\thanks{Computer Science and Artificial Intelligence Laboratory (CSAIL) at MIT, Cambridge, MA.  {\tt tasos@theory.csail.mit.edu}. This work was done while visiting Google, Inc., New York, NY.}
\and Cliff Stein\thanks{Department of IEOR, Columbia University.  {\tt cliff@ieor.columbia.edu}.  This work was done while visiting Google, Inc., New York, NY.}
\and Zoya Svitkina\thanks{Department of Computer Science, Cornell University. {\tt zoya@cs.cornell.edu}.  This work was done while visiting Google, Inc., New York, NY.}
}

\begin{document}

\maketitle
\thispagestyle{empty}
\begin{abstract}
An existing approach for dealing with massive data sets is to stream
over the input in few passes and perform computations with sublinear
resources.  This method does not work for truly massive data where
even making a single pass over the data with a processor is
prohibitive. Successful log processing systems in practice such as
Google's MapReduce and Apache's Hadoop use multiple machines. They
efficiently perform a certain class of highly distributable
computations defined by local computations that can be applied in any
order to the input.

Motivated by the success of these systems, we introduce a simple
algorithmic model for massive, unordered, distributed (mud)
computation. We initiate the study of understanding its computational
complexity.  Our main result is a positive one: any unordered function
that can be computed by a streaming algorithm can also be computed
with a mud algorithm, with comparable space and communication
complexity.  We extend this result to some useful classes of
approximate and randomized streaming algorithms.  We also give
negative results, using communication complexity arguments to prove
that extensions to private randomness, promise problems and
indeterminate functions are impossible.

We believe that the line of research we introduce in this paper has
the potential for tremendous impact.  The distributed systems that
motivate our work successfully process data at an unprecedented scale,
distributed over hundreds or even thousands of machines, and perform
hundreds of such analyses each day.  The mud model (and its
generalizations) inspire a set of complexity-theoretic questions that
lie at their heart.
\end{abstract}

\newpage

\section{Introduction}

We now have truly massive data sets, many of which are
generated by logging events in physical systems.
For example, 
data sources such as IP traffic logs, web page repositories,
search query logs, retail and financial transactions, and other sources 
consist of billions of items per day, and are accumulated over many
days.  Internet search companies such as Google, Yahoo!, and MSN,
financial companies such as Bloomberg, retail businesses such
as Amazon and WalMart, and other companies use this type of
data.  

In theory, we have formulated the {\em data stream model} to study
algorithms that process such truly massive data sets.  Data stream
models~\cite{HRR,AMS} make one pass over the logs, read and process
each item on the stream rapidly and use local storage of size
sublinear---typically, polylogarithmic---in the input.  There is now a
large body of algorithms and lower bounds in data stream models
(see~\cite{muthu} for a survey).  

Yet, streaming models alone are not sufficient.  For example, logs of
Internet activity are so large that no single processor can make even
a single pass over the data in a reasonable amount of time.  
The solution in practice has been to deploy more machines, distribute the
data over these machines and process different pieces of data in
parallel.  
For example, Google's
MapReduce~\cite{mapreduce} and Apache's Hadoop~\cite{hadoop} are
successful large scale distributed platforms that can process many
terabytes of data at a time, distributed over hundreds or even
thousands of machines, and process hundreds of such analyses each day.
A reason for their success is that logs-processing
algorithms written for these platforms have a simple form that let
the platform process the input in an arbitrary order, and
combine partial computations using whatever communication pattern is convenient.  

In this paper, we introduce a simple model for these algorithms, which
we refer to as ``mud'' (Massive, Unordered, Distributed) algorithms.
This computational model raises several interesting complexity
questions which we address.  Almost all the work in streaming
including the seminal~\cite{HRR,AMS} and its extensions~\cite{ruhl}
have been motivated by massive data computations, making one or more
linear passes over the data. The algorithms developed in this area
have in many cases found applications to distributed data processing,
e.g., motivated by sensor networks.  Our work is the first to address
this distributed model specifically, and attempt to understand its
power and limitations.


\subsection{Mud algorithms}

Distributed platforms like MapReduce and Hadoop are engines 
for executing arbitrary tasks with a certain simple structure over
many machines.  These platforms can solve many different kinds of
problems, and in particular are used extensively for analyzing logs.
Logs analysis algorithms written for these platforms consist of three
functions: (1) a {\em local} function to take a single input data item and
output a message, (2) an {\em aggregation} function to combine pairs of
messages, and in some cases (3) a final post-processing step.  The
distributed platform assumes that the local function can be applied to
the input data items independently in parallel, and that the aggregation function
can be applied to pairs of messages in any order.  This allows the
platform to synchronize the machines very coarsely (assigning them to
work on whatever chunk of data becomes available), 
 and avoids the need for machines to share vast amounts of
data (thereby eliminating communication bottlenecks)---yielding a
highly distributed, robust execution in practice.

\medskip
\noindent
{\bf Example.} Consider this simple logs analysis algorithm to compute the
sum of squares of a large set of numbers:\footnote{This is 
expressed in written Sawzall~\cite{saw} language, a language at Google
for logs processing,
that runs on the MapReduce
platform.  The example is a complete
Sawzall program minus some type declarations.}
\begin{quote}
\begin{quote}
\begin{quote}
{\small
\begin{verbatim}
x = input_record;      
x_squared = x * x;
aggregator: table sum;  
emit aggregator <- x_squared;
\end{verbatim}
}
\end{quote}
\end{quote}
\end{quote}
This program is written as if it only runs on a single input
record, since it is interpreted as the local function in MapReduce.
Instantiating the {\tt aggregator} object as a ``table'' of type
``sum'' signals MapReduce to use summation as its aggregation
function.  ``Emitting'' {\tt x\_squared} into the aggregator defines
the message output by the local function.
When MapReduce executes this program, the final output is the result
of aggregating all the messages (in this case the sum of the
squares of the numbers).  This can then be post-processed in some way
(e.g., taking the square root, for computing the $L_2$ norm).
Large numbers of algorithms of this form are used daily for processing logs~\cite{saw}.

\paragraph{Definition of a mud algorithm.} 
We now formally define a {\em mud algorithm} as a triple $m = (\fil,
\oplus, \cleanup)$.  The local function $\fil: \X \to \states$ maps an
input item to a message, the aggregator $\oplus:
\states \times \states \to \states$ maps two messages to a
single message, and the post-processing operator $\cleanup :
\states \to \X$ produces the final output.  
The output can depend on the
order in which $\oplus$ is applied.  Formally, let $\tree$ be an arbitrary
binary tree circuit with $n$ leaves.  We use $m_\tree(\seq)$ to denote
the $q \in \states$ that results from applying $\oplus$ to the 
sequence $\fil(x_1), \dots, \fil(x_n)$ along the 
topology of $\tree$ with an arbitrary permutation of these inputs as its leaves.
The overall output of the mud algorithm is then
$\cleanup(m_\tree(\seq))$, which is a function $\X^n \to \X$.  Notice
that $\tree$ is {\em not} part of the algorithm definition, but
rather, the algorithm designer needs to make sure that
$\cleanup(m_\tree(\seq))$ is independent of $\tree$.\footnote{This is
implied if $\oplus$ is associative and commutative; however, this is
not necessary.} We say that a mud algorithm {\em computes} a function
$f$ if $\cleanup(m_\tree(\cdot)) = f$ for all trees $\tree$.

We give two examples. On the left is a mud algorithm to compute the
total span ($\max - \min$) of a set of integers.  On the right is a mud
algorithm to compute a uniform random sample of the {\em unique} items
in a set (i.e, items that appear at least once) by using an
approximate minwise hash function $h$ (see~\cite{broder,datar} for
details):

\medskip
\begin{tabular}{|l|l|}\hline
$\fil(x) = \langle x, x \rangle$ & $\fil(x) = \langle x, h(x), 1
  \rangle$ \\ & \\
$\oplus( \langle a_1, b_1 \rangle , \langle a_2, b_2 \rangle ) = \langle \min(a_1, a_2), \max(b_1, b_2) \rangle$ &
$\oplus( \langle a_1, h(a_1), c_1 \rangle , \langle a_2, h(a_2), c_2 \rangle ) $ \\
& \;\;\; = $ \left\{ \begin{array}{ll}
  \langle a_i, h(a_i), c_i \rangle &  {\rm if} \; h(a_i) < h(a_j)\\
  \langle a_1, h(a_1), c_1+c_2 \rangle &  {\rm otherwise}
\end{array} \right. $ \\ & \\
$\cleanup( \langle a, b \rangle ) = b - a$ & $\cleanup(\langle a, b, c \rangle) = a \; {\rm if} \; c = 1$ \\ \hline
\end{tabular}
\medskip


\junk{
\begin{eqnarray*}
\fil(x) & = & \langle x, x \rangle \\
\oplus( \langle a_1, b_1 \rangle , \langle a_2, b_2 \rangle ) & = & \langle \min(a_1, a_2), \max(b_1, b_2) \rangle \\
\cleanup( \langle a, b \rangle ) & = & b - a
\end{eqnarray*}
}

The communication complexity of a mud 
algorithm is $\log |Q|$, the number of bits needed to represent a
``message'' from one component to the next.
We consider the \{space, time\} complexity of a mud algorithm to be the maximum \{space, time\} complexity
of its component functions.\footnote{This is the only thing that is under the
control of the algorithm designer; indeed the actual execution
time---which we do not formally define here---will be a function of
the number of machines available, runtime behavior of the platform and these local
complexities.}

\subsection{How complex are mud algorithms?}

We wish to understand the complexity of mud algorithms.  Recall that a
mud algorithm to compute a function must work for all computation
trees over $\oplus$ operations; now consider the following tree:
$\oplus(\oplus(\ldots\oplus(\oplus(q, \fil(x_1)),\fil(x_2)),\dots,\fil(x_{k-1})),\fil(x_k))$.
This sequential application of $\oplus$ corresponds to the 
conventional {\em streaming} model (see eg. survey of~\cite{muthu}).  

Formally, a streaming algorithm is given by $s = (\sigma, \cleanup)$,
where $\sigma : \states \times \X \to \states$ is an operator applied
repeatedly to the input stream, and $ \cleanup: \states \to \X$
converts the final state to the output. The notation $s^q(\seq)$
denotes the state of the streaming algorithm after starting at state
$q$, and operating on the sequence $\seq = x_1, \dots, x_k$ {\em in
that order}, that is, $ s^q(\seq) =
\sigma(\sigma(\ldots\sigma(\sigma(q,x_1),x_2),\dots,x_{k-1}),x_k) $.
On input $\seq \in \X^n$, the streaming algorithm computes
$\cleanup(s^0(\seq))$, where $0$ is the starting state.  As in mud, we
define the communication complexity to be $\log |Q|$ (which is
typically polylogarithmic), and the \{space, time\} complexity as the
maximum \{space, time\} complexity of $\sigma$ and $\cleanup$.

Streaming algorithms can compute whatever mud algorithms can compute: 
given a mud algorithm $m=(\fil, \oplus, \cleanup)$,
there is a streaming algorithm $s = (\sigma, \cleanup)$ of the same complexity 
with same output, by setting
$\sigma(\state, x) = \oplus(\state, \fil(x))$.
The central question then is,  {\em can a mud algorithm compute
whatever a streaming algorithm computes? }
It is immediate that there are streaming computations that
cannot be simulated by mud algorithms.  For example, consider a streaming
algorithm that counts the number of occurrences of the {\em first} element
in the stream: no mud algorithm can accomplish this since it cannot determine the 
first element in the input. Therefore, in order to be fair, since mud algorithms work on unordered data,
we restrict our attention to functions $\X^n \to \X$ that are {\em
symmetric} (order-invariant) and address this central question. 

\subsection{Our Results}

We present the following positive and negative results comparing mud to streaming algorithms,
restricted to symmetric functions:
\begin{itemize}
\item We show that any deterministic streaming algorithm that computes a symmetric
function $\X^n \to \X$ can be simulated by a mud algorithm with the
same communication complexity, and the square of its space complexity.
This result generalizes to certain approximation
algorithms, and randomized algorithms with public randomness.

\item
We show that the claim above does not extent to richer symmetric function classes, 
such as when the function comes with a {\em promise} that the
domain is guaranteed to satisfy some property (e.g., finding the
diameter of a graph known to be connected), or the function is {\em indeterminate}, i.e., 
one of many possible outputs is allowed for ``successful
computation.'' (e.g., finding a number in the highest 10\% of a set of
numbers.)  Likewise, with private randomness, the claim above is no longer true. 
\end{itemize}

\junk{
In both our positive and negative results we consider streaming and
$\mud$ algorithms with comparable space resources---roughly, $\log
|Q|$, which plays an important role in the application of
communication complexity arguments.  Consistent with our motivation
for looking at problems on massive data sets, and for clarity of
presentation, we will fix $Q$ throughout the paper as an alphabet with
$\log |Q| = \polylog n$, and restrict our attention to algorithms
bounded by space $\polylog n$.  However, many of our results
generalize to wider ranges of alphabet sizes and space complexity,
which we will point out when appropriate.
}
   

The simulation in our result
takes time $\Omega(2^{\polylog (n)})$ from the use of Savitch's
theorem. So while not a practical algorithm, our result implies that
if we wanted to separate mud algorithms from streaming algorithms for symmetric
functions, we
need techniques other than communication complexity-based arguments. 

Also, when we consider symmetric problems that have been addressed in the
streaming literature, they seem to always yield mud algorithms
(e.g., all streaming algorithms that allow insertions and deletions in
the stream, or are based on various {\em sketches}~\cite{AMS} can be
seen as mud algorithms).  
In fact, we are not aware of a specific
problem that has a streaming solution, but no mud algorithm with
comparable complexity (up to polylog factors in space and per-item
time).\footnote{There are specific algorithms---such as one of the
algorithms for estimating $F_2$ in~\cite{AMS}---that are sequential
and not mud algorithms, but there are other alternative mud algorithms
with similar bounds for the problems they solve.}  Our result here
provides some insight into this intuitive state of our knowledge and
presents rich function classes for which distributed streaming (mud)
is provably as powerful as sequential streaming.
%
%

\subsection{Techniques}

One of the core arguments used to prove our positive results comes
from an observation in communication complexity.  
Consider evaluating a symmetric function $f(\seq)$ given two disjoint portions
of the input $\seq = \seq_A \cdot \seq_B$, in each of the two following models.  In the {\em
one-way communication model} (OCM), David knows portion $\seq_A$, and
sends a single message $D(\seq_A)$ to Emily who knows portion
$\seq_B$; she then outputs $E(D(\seq_A), \seq_B) = f(\seq_A \cdot
\seq_B)$. In the {\em simultaneous communication model} (SCM) both
Alice and Bob send a message $A(\seq_A)$ and $B(\seq_B)$ respectively,
simultaneously to Carol who must compute $f(\seq_A \cdot \seq_B)$.
Clearly, OCM protocols can simulate SCM protocols.\footnote{The SCM here is identical to the
simultaneous message model~\cite{babai} or oblivious communication
model~\cite{pavel} studied previously if there are $k=2$ players. For
$k>2$, our mud model is not the same as in previous
work~\cite{babai,pavel}.  The results in~\cite{babai,pavel} as it
applies to us are not directly relevant since they only show examples
of functions that separate SCM and OCM significantly.}  
At the core, our result relies on observing that SCM protocols can simulate 
OCMs too, for symmetric functions $f$, by guessing the inputs that result in
the particular message received by a party.
\junk{
Is SCM significantly
weaker than OCM?This is a
simplification of our question concerning mud vs. streaming.

We can simulate an OCM process in the SCM as follows. Both Alice and
Bob imagine that they are David, and send messages $D(\seq_A)$ and
$D(\seq_B)$.  Carol would like to compute $E( D(\seq_A), \seq_B)$, but
she only has $D(\seq_B)$, and not $\seq_B$ itself.  So, she guesses a
sequence $\seq'_B$ such that $\seq'_B = D(\seq_B)$, and outputs $E(
D(\seq_A), \seq'_B)$.  Using the symmetry of $f$, it can be
shown that Carol outputs $f(\seq_A \cdot \seq_B)$ correctly: 
$E( D(\seq_A), \seq'_B) = f(\seq_A \cdot \seq'_B) = f(\seq'_B \cdot \seq_A) = E(
D(\seq_B), \seq_A) = f(\seq_B \cdot \seq_A) = f(\seq_A \cdot \seq_B).$
}

To prove our main result---that mud can simulate streaming---we apply
the above argument many times over an arbitrary tree topology of
$\oplus$ computations, using Savitch's theorem to guess input
sequences that match input states of streaming computations.  This is
delicate because we can use the symmetry of $f$ only at the root of
the tree; simply iterating the argument at every node in the
computation tree independently would yield weaker results that would
force the function to be symmetric on {\em subsets} of the input,
which is not assumed by our theorem.

To prove our negative results, we also use communication
limitations---of the intermediate SCM.  We define order-independent problems easily solved by
a single-pass streaming algorithm and then formulate instances that
require a polynomial
amount of communication in the SCM. The order-independent problems
we create are variants of parity and index problems that are traditionally used
in communication complexity lower bounds. 

\section{Main Result}
\label{sec:positive}
In this section we give our main result, that any symmetric function
computed by a streaming algorithm can also be computed by a mud
algorithm. 

\subsection{Preliminaries}

 
As is standard,  we fix the space and
communication to be $\polylog(n)$.\footnote{The results in this paper
extend to other sub-linear (say $\sqrt{n}$) space, and communication bounds in a natural
way.}

\begin{definition}
A symmetric function $f: \Xn \to \X$ is in the class $\mud$ if there
exists a $\polylog(n)$-communication, $\polylog(n)$-space mud algorithm $m = (\fil, \oplus, \cleanup)$ such that for all
$\seq \in \X^n$, and all computation trees $\tree$, we have
$\cleanup(m_\tree(\seq)) = f(\seq)$.
\end{definition}


\begin{definition}
A symmetric function $f: \Xn \to \X$ is in the class $\stream$ if
there exists a $\polylog(n)$-communication, $\polylog(n)$-space streaming algorithm $s=(\sigma, \cleanup)$ such that for
all $\seq \in \X^n$ we have $\cleanup(s^0(\seq)) = f(\seq)$.
\end{definition}

Note that for subsequences $\seq_{\alpha}$ and $\seq_{\beta}$, we get
$s^q(\seq_{\alpha} \cdot \seq_{\beta}) = s^{s^q(\seq_{\alpha})}(\seq_{\beta}).$
%
We can apply this identity to obtain the following simple lemma.
\begin{lemma}
\label{lem:switchab}
Let $\seq_{\alpha}$ and $\seq'_{\alpha}$ be two strings and $q$ a
state such that
$s^q(\seq_{\alpha}) = s^q(\seq'_{\alpha})$.  Then for any string
$\seq_{\beta}$, we have 
$s^q(\seq_{\alpha} \cdot \seq_{\beta}) = s^q(\seq'_{\alpha}\cdot
  \seq_{\beta})$.
\end{lemma}
\begin{proof}
We have $s^q(\seq_{\alpha} \cdot \seq_{\beta}) = 
s^{s^q(\seq_{\alpha})}(\seq_{\beta}) = 
s^{s^q(\seq'_{\alpha})}(\seq_{\beta}) = 
s^q(\seq'_{\alpha}\cdot
  \seq_{\beta})
$
\end{proof}
Also, note that for some $f \in \stream$, because $f$ is symmetric, the output $\cleanup(s^0(\seq))$ of a
streaming algorithm $s = (\sigma, \cleanup)$ that computes it must be
invariant over all permutations of the input; i.e.:
\begin{eqnarray}
\label{fact:pi}
\forall x \in \X^n, \text{ permutations } \pi:~~~\cleanup(s^0(\seq)) = f(\seq) = f(\pi(\seq)) = \cleanup(s^0(\pi(\seq)))  
\end{eqnarray}
This fact about the {\em output} of $s$ does not necessarily
mean that the {\em state} of $s$ is permutation-invariant; indeed,
consider a streaming algorithm to compute the sum of $n$ numbers that
for some reason remembers the first element it sees (which is
ultimately ignored by the function~$\cleanup$).  In this case the
state of $s$ depends on the order of the input, but the final output
does not.

\subsection{Statement of the result}

We argued that streaming algorithms can simulate mud algorithms by
setting $\sigma(\state,x) = \oplus(\fil(x),x)$, which implies
$\mud \subseteq \stream$.  The main result in this paper is:

\begin{theorem}
\label{thm:main}
For any symmetric function $f: \X^n \to \X$  computed by a
$g(n)$-space, $c(n)$-communication streaming algorithm $(\sigma, \cleanup)$, with $g(n) =
\Omega(\log n)$ and $c(n) = \Omega(\log n)$, there exists a  $O(c(n))$-communication, $O(g^2(n))$-space mud algorithm
$(\fil, \oplus, \cleanup)$ that also computes $f$.
\end{theorem}

\noindent This immediately gives: $\mud=\stream$.


\subsection{Proof of Theorem~\ref{thm:main}}
\label{sec:proof}

We prove Theorem~\ref{thm:main} by simulating an arbitrary streaming
algorithm with a mud algorithm.  
The main challenges of the simulation
are in 

\medskip

{\bf (i)} achieving polylog communication complexity in the messages sent between $\oplus$ operations,

\smallskip
{\bf (ii)} achieving polylog space complexity for computations needed to support the protocol above, and 

\smallskip
{\bf (iii)} extending the methods above to work for an arbitrary computation tree. 

\medskip

\noindent We tackle these three challenges in order with the full proof given later.

\paragraph{Communication complexity.}
Consider the final application of $\oplus$ (at the root of the tree $\tree$) in a mud computation.  The inputs to this function are two
messages $q_A, q_B \in \states$ that are computed independently from
a partition $\seq_A, \seq_B$ of the input.  The output is a state
$q_C$ that will lead directly to the overall output $\cleanup(q_C)$.
This is similar to the task Carol faces in SCM:  the input
$\X^n$ is split arbitrarily between Alice and Bob, who
 independently process their input (using unbounded computational resources), but then must transmit only a single symbol from $\states$ to Carol;
Carol then performs some final processing (again, unbounded),
and outputs an answer in $\X$.  We show:


\begin{theorem}
\label{thm:3cm}
Every function $f \in \stream$ can be computed in the SCM with communication $\polylog(n)$.
\end{theorem}


\begin{proof} 
Let $s = (\sigma, \cleanup)$ be a streaming algorithm that computes $f$.
We assume (wlog) that the streaming algorithm $s$
maintains a counter in its state $q \in Q$ indicating the number of
input elements it has seen so
far.

We compute $f$ in the SCM as follows.
Let $\seq_A$ and $\seq_B$ be the partitions of the input sequence $\seq$
sent to Alice and Bob.
Alice simply runs the streaming algorithm on her input sequence to
produce the state $q_A = s^0(\seq_A)$, and sends this to Carol.
Similarly, Bob sends $q_B = s^0(\seq_B)$ to Carol.  
Carol receives the states $q_A$ and $q_B$, which contain the sizes
$n_A$ and $n_B$ of the input sequences $\seq_A$ and $\seq_B$.  
She then finds sequences $\seq'_A$ and $\seq'_B$ of length $n_A$ and $n_B$
such that $q_A = s^0(\seq'_A)$ and $q_B = s^0(\seq'_B)$.  (Such sequences
must exist since $\seq_A$ and $\seq_B$ are candidates.)
Carol then outputs $\cleanup(s^0(\seq'_A \cdot \seq'_B))$.
To complete the proof: 
$$
\begin{array}{rcll}
\cleanup(s^0(\seq'_A \cdot \seq'_B))
& = & \cleanup(s^0(\seq_A \cdot \seq'_B)) & (\mbox{by Lemma}~\ref{lem:switchab}) \\
& = & \cleanup(s^0(\seq'_B \cdot  \seq_A)) & (\mbox{by }\eqref{fact:pi}) \\
& = & \cleanup(s^0(\seq_B \cdot  \seq_A)) & (\mbox{by 
Lemma}~\ref{lem:switchab}) \\
& = & \cleanup(s^0(\seq_A \cdot  \seq_B)) & (\mbox{by }\eqref{fact:pi}) \\
& = & f(\seq_A \cdot \seq_B) & (\mbox{by the correctness of $s$}) \\
& = & f(\seq) . &
\end{array}
$$

\vspace{-.3in}
\end{proof}

\paragraph{Space complexity.}
The simulation above uses space linear in the input. 
We now give a more space-efficient implementation
of Carol's computation.
More precisely, if the streaming algorithm uses space $g(n)$, we show
how Carol can use only space $O(g^2(n))$; this space-efficient
simulation will eventually be the
algorithm used by $\oplus$ in our mud algorithm.

\begin{lemma}
\label{lemma:savitch}
Let $s = (\sigma, \cleanup)$ be a $g(n)$-space streaming algorithm
with $g(n) = \Omega(\log n)$.  Then, there is a $O(g^2(n))$-space
algorithm that, given states $q_A, q_B \in \states$ and lengths $n_A, n_B \in [n]$, outputs a state
$q_C = s^0(\seq_C)$, where $\seq_C = \seq'_A \cdot \seq'_B$ for some
$\seq'_A, \seq'_B$ of lengths $n_A, n_B$ such that $s^0(\seq'_A) =
q_A$ and $s^0(\seq'_B) = q_B$. (If such a $q_C$ exists.)
\end{lemma}

\begin{proof}
Note that there may be many $\seq'_A, \seq'_B$ that
satisfy the conditions of the theorem, and thus there are many valid answers for $q_C$.  We
only require an arbitrary such value.
However, if we only have $g^2(n)$ space, and $g^2(n)$ is sublinear, we cannot even
write down $\seq'_A$ and $\seq'_B$.  Thus we need to be careful about how we find $q_C$.

Consider a non-deterministic algorithm for computing
a valid $q_C$.  
First, guess the symbols of $\seq'_A$ one at a time, simulating the
streaming algorithm $s^0(\seq'_A)$ on the guess.
If after $n_A$ guessed symbols we have $s^0(\seq'_A) \neq q_A$, reject
this branch.
Then, guess the symbols of $\seq'_B$, simulating (in parallel)
$s^0(\seq'_B)$ and $s^{q_A}(\seq'_B)$.
If after $n_B$ steps we have $s^0(\seq'_B) \neq q_B$, reject this
branch; otherwise, output $q_C = s^{q_A}(\seq'_B)$.
This is a non-deterministic, $O(g(n))$-space algorithm for
computing a valid $q_C$.  By Savitch's theorem~\cite{savitch}, it follows that
$q_C$ can be computed by a deterministic, $g^2(n)$-space
algorithm. (The application of
Savitch's theorem in this context amounts to a dynamic program for
finding a state $q_C$ such that the streaming algorithm can get from
state $q_A$ to $q_C$ and from state $0$ to $q_B$ using the same input
string of length $n_B$.)
\end{proof}

The running time of this algorithm is super-polynomial from
the use of Savitch's theorem, which dominates the 
running time in our simulation.


\paragraph{Finishing the proof for arbitrary computation trees.}
To prove Theorem~\ref{thm:main}, we will simulate an arbitrary streaming algorithm with a mud algorithm, setting $\oplus$ 
to Carol's procedure, as implemented in Lemma~\ref{lemma:savitch}.  The remaining challenge
is to show that the computation is successful on an arbitrary computation
tree;  we do this by relying on the symmetry of $f$ and the correctness of Carol's procedure.


\newcommand{\rseq}{\seq^*}
\newcommand{\rx}{x^*}
\newcommand{\hseq}{\hat{\seq}}

\medskip
\noindent {\em Proof of Theorem~\ref{thm:main}:}
Let $f \in \stream$ and let $s = (\sigma, \cleanup)$ be a streaming
algorithm that computes $f$.  We assume wlog that $s$ includes in its state
$q$ the number of inputs it has seen so far.
We define a mud algorithm $m = (\fil, \oplus, \cleanup)$ where
$\fil(x) = \sigma(0,x)$, and using the same $\cleanup$ function as $s$ uses.  
The function $\oplus$, given $q_A, q_B \in \states$ and input sizes
$n_A, n_B$, outputs some $q_C = q_A \oplus q_B = s^0(\seq_C)$ 
as in Lemma~\ref{lemma:savitch}. 
To show the correctness of $m$, we need to show
that $\cleanup(m_\tree(\seq)) = f(\seq)$ for all computation trees
$\tree$ and all $\seq \in \X^n$.  For the remainder of the proof, let
$\tree$ and $\rseq = (\rx_1, \dots, \rx_n)$ be an arbitrary tree and input sequence, respectively.
The tree $\tree$ is a binary in-tree with $n$ leaves.  Each node
$v$ in the tree outputs a state $q_v \in \states$, including the
leaves, which output a state $q_i = \fil(\rx_i) = \sigma(0,\rx_i) =
s^0(\rx_i)$.  
The root $r$ outputs $q_r$, and so we need to prove that
$\cleanup(q_r) = f(\rseq)$.

The proof is inductive.
We associate with each node $v$ a ``guess sequence,'' $\seq_v$ which for internal
nodes is the sequence $\seq_C$ as in Lemma~\ref{lemma:savitch}, and for
leaves $i$ is the single symbol $\rx_i$. Note that for all nodes $v$, we have $q_v = s^0(\seq_v)$, and 
the length of $\seq_v$ is
equal to the number of leaves in the subtree rooted at $v$.
Define a {\em frontier} of tree nodes to be a set of nodes such that
each leaf has exactly one ancestor in the set. (A node is considered
an ancestor of itself.)  The root itself is a frontier, as is the
complete set of leaves.
We say a frontier $V = \{ v_1, \dots, v_k\}$ is {\em correct} if the
streaming algorithm on the data associated with the frontier is
correct, that is, 
$\cleanup(s^0(\seq_{v_1} \cdot \seq_{v_2} \cdot \dots \cdot \seq_{v_k})) = f(\rseq).$
Since the guess sequences of a frontier always have total length $n$,
the correctness of a frontier set is invariant of how the set is
ordered (by ~\eqref{fact:pi}).
Note that the frontier set consisting of all leaves is immediately correct
by the correctness of $f$.
The correctness of our mud algorithm would follow from the
correctness of the root as a frontier set, since at the
root, correctness implies $\cleanup(s^0(\seq_r)) = \cleanup(q_r)
= f(\rseq)$.

To prove that the root is a correct frontier, it suffices to define an
operation to take an arbitrary correct frontier $V$ with at least two
nodes, and produces another correct frontier $V'$ with one fewer node.
We can then apply this
operation repeatedly until the unique frontier of size one (the root)
is obtained. 
Let $V$ be an arbitrary correct frontier with at least two nodes.  We
claim that $V$ must contain two children $a, b$ of the same node
$c$.\footnote{Proof: consider one of the nodes $a \in V$ furthest
from the root.  Suppose its sibling $b$ is not in $V$.  Then any leaf in
the tree rooted at $b$ must have its ancestor in $V$ further from $r$ than
$a$; otherwise a leaf in the tree rooted at $a$ would have two
ancestors in $V$.  This contradicts $a$ being furthest from the
root.}  
To obtain $V'$ we replace $a$ and $b$ by their parent $c$.
Clearly $V'$ is a frontier, and so it remains to show that $V'$ is correct.
We can write $V$ as $\{ a, b, v_1, \dots, v_k \}$, and so 
$V' = \{c, v_1, \dots, v_k \}$.  
For ease of notation, let $\hseq = \seq_{v_1} \cdot \seq_{v_2} \cdot \dots \cdot \seq_{v_k}$.

The remainder of the argument follows the logic in the proof of
Theorem~\ref{thm:3cm}.  Observe that we now have to be careful that
the guess for a string is the same length as the original string; this
property is guaranteed in Lemma~\ref{lemma:savitch}.
$$
\begin{array}{rcllr}
f(\rseq) 
& = & \cleanup(s^0(\seq_a \cdot \seq_b \cdot \hseq)) 
& (\mbox{by the   correctness of } V) & \\
& = & \cleanup(s^0(\seq'_a \cdot \seq_b \cdot \hseq))  
& (\mbox{by   Lemma}~\ref{lem:switchab}) & \\
& = & \cleanup(s^0(\seq_a \cdot \seq'_b \cdot \hseq)) 
& (\mbox{by }\eqref{fact:pi}) & \\
& = & \cleanup(s^0(\seq'_b \cdot \seq'_a \cdot \hseq)) 
& (\mbox{by   Lemma}~\ref{lem:switchab}) & \\
& = & \cleanup(s^0(\seq'_a \cdot \seq'_b \cdot \hseq)) 
& (\mbox{by}  ~\eqref{fact:pi}) & \\
& = & \cleanup(s^0(\seq_c \cdot \hseq)) 
& (\mbox{by   Lemma}~\ref{lemma:savitch}) & \square \end{array}
$$

\subsection{Extensions to randomized and approximation algorithms}

We have proved that any deterministic streaming computation of a
symmetric function can be simulated by a mud algorithm.  However most
nontrivial streaming algorithms in the literature rely on randomness,
and/or are approximations.  Still, our results have interesting
implications as described below.

Many streaming algorithms for approximating a function $f$ work by computing
some other function $g$ exactly over the stream, and from that obtaining an approximation
$\tilde{f}$ to $f$, in postprocessing.  
For example, sketch-based streaming algorithms maintain counters computed by inner products $c_i =
\langle \seq, \bvec{v}_i \rangle$ where $\seq$ is the input vector and each $\bvec{v}_i$
is some vector chosen by the algorithm.
From the set of $c_i$'s, the
algorithms compute $\tilde{f}$. 
As long as 
$g$ is a symmetric function (such as the counters), our simulation results apply to $g$ and hence to
the approximation of $f$:  such streaming algorithms, approximate
though they are, have equivalent mud algorithms. 
This is a strengthening of Theorem~\ref{thm:main} to approximations.




Our discussion above can be
formalized easily for deterministic algorithms. 
There are however some details in formalizing it 
for randomized algorithms.
Informally, we focus on the class of randomized streaming algorithms that are order-independent for
particular choices of random bits, such as  all the randomized sketch-based~\cite{AMS,I}
streaming algorithms.
Formally, 

\begin{definition}
\label{def:rss}
A symmetric function $f: \Xn \to \X$ is in the class $\rstream$ if 
there exists a set of $\polylog(n)$-communication, $\polylog(n)$-space 
streaming algorithms $\{ s^R=(\sigma^R, \cleanup^R) \}_{R \in \{0,1\}^k}$,
$k = \polylog(n)$, such that for all $\seq \in X^n$, 
\begin{enumerate}
\item $\Pr_{R \sim \set{0,1}^k }
  \left[ \cleanup^R(s^R(\seq)) = f(\seq) \right] \geq \frac{2}{3}$, and
\item for all $R \in \set{0,1}^k$, and permutations
  $\pi$, $\cleanup^R(s^R(\seq)) = \cleanup^R(s^R(\pi(\seq)).$ 
\end{enumerate}
\end{definition}

\noindent 
We define the randomized variant of $\mud$ analogously.


\begin{definition}
\label{def:rmud}
A symmetric function $f: \Xn \to \X$ is in $\rmud$ if 
there exists a set of $\polylog(n)$-communication, $\polylog(n)$-space 
mud algorithms $\{ m^R=(\fil^R, \oplus^R, \cleanup^R) \}_{R \in \{0,1\}^k}$,
$k = \polylog(n)$, such that for all $\seq \in X^n$, 
\begin{enumerate}
\item 
for all computation trees $\tree$, we have $\Pr_{R \sim \set{0,1}^k }   
\left[ \cleanup^R(m^R_\tree(\seq)) =  f(\seq) \right]     \geq \frac{2}{3}$, and
\item for all $R \in \set{0,1}^k$, permutations $\pi$, and pairs
of trees $\tree, \tree'$, we have
$\cleanup^R(m^R_\tree(\seq)) = \cleanup^R(m^R_{\tree'}(\pi(\seq)))$.
\end{enumerate}
\end{definition}

\noindent The second property in each of the definitions ensures that each
particular algorithm ($s^R$ or $m^R$) computes a deterministic
symmetric function after $R$ is chosen.  This makes it straightforward
to extend Theorem~\ref{thm:main} to show
$\rmud = \rstream$.

\section{Negative Results}

In the previous section, we demonstrated conditions under which mud
computations can simulate streaming computations.  We saw, explicitly
or implicitly, that we have mud algorithms for a function 

\bigskip

{\bf (i)} that is total, ie., defined on all inputs, 

\medskip

{\bf (ii)} that has one unique output value, and, 

\medskip

{\bf (iii)} that has a streaming algorithm that, if randomized, uses public randomness.  

\bigskip

In this section, we show that each one of these
conditions is necessary: if we drop any of them, we can separate mud from streaming.
Our separations are based on communication complexity lower bounds 
in the SCM model, which suffices (see the ``communication complexity'' paragraph in Section~\ref{sec:proof}). 

\junk{
The precise definitions appear in Sections \ref{sec:promise}, but we state a preliminary observation that
justifies our use of the SCM model.

\begin{proposition}
If a symmetric function $f : \X^n \to \X$ is in $\mud$, $\imud$ or $\pmud$ then it can also be computed in the SCM.
\end{proposition}

\begin{proof}
Suppose $\bvec{x} \in \X^n$ is split into $\bvec{x}_A$ and
$\bvec{x}_B$ and given to Alice and Bob.  Alice computes
$\tree^A_{\fil,\oplus}(\bvec{X}_A)$ for some arbitrary computation
tree $\tree^A$ and sends it to Carol; Bob computes
$\tree^B_{\fil,\oplus}(\bvec{X}_A)$ for some $\tree^B$ and
sends it to Carol; finally, Carol can compute
$\cleanup(\tree_{\fil,\oplus}(\bvec{x}))$ for some tree $\tree$.
\end{proof}
}

\subsection{Private Randomness}

In the definition of $\rmud$, we assumed that the same $R$ was given to each component; i.e, public randomness.
We show that this is necessary in order
to simulate a randomized streaming algorithm, even for the
case of total functions.  
Formally, we prove:

\begin{theorem}\label{thm:private_randomness}
There exists a symmetric total function $f\in \rstream$, such that there is no randomized mud algorithm for computing $f$ using only private randomness.
\end{theorem}

In order to prove Theorem~\ref{thm:private_randomness}, we demonstrate a total
function $f$ that is computable by a single-pass, randomized
$\mbox{polylog}(n)$-space streaming algorithm, but any SCM 
protocol for $f$ with private randomness has communication complexity
$\Omega(\sqrt{n})$.  Our proof uses a reduction from the
{\em string-equality problem} to a problem that we call {\sc SetParity}.
In the later problem, we are given a collection of records
$
S  =  (i_1, b_1), (i_2, b_2), \ldots, (i_n, b_n),
$
where for each $j\in [n]$, we have $i_j\in \{0,\ldots,n-1\}$, and $b_j\in \{0,1\}$.
We are asked to compute the following function, which is clearly a total function under a natural encoding of the input:

{\small \begin{eqnarray*}
f(S) & = & \left\{\begin{array}{ll}
1 & \mbox{ if } \forall t\in\{0,\ldots,n-1\}, \sum_{j:i_j=t} b_j \mbox{ mod } 2 = 0\\
0 & \mbox{ otherwise}
\end{array}\right.
\end{eqnarray*}}
We give a randomized streaming algorithm that computes $f$ using the
$\epsilon$-biased generators of~\cite{NN}.  Next, in order to
lower-bound the communication complexity of a SCM protocol for {\sc
SetParity}, we use the fact that any SCM protocol for string-equality
has complexity $\Omega(\sqrt{n})$\cite{stringeq1,stringeq2}.  Due to lack of
space, the remainder of the proof of
Theorem~\ref{thm:private_randomness} is given in the appendix.

\subsection{Promise Functions}
\label{sec:promise}

In many cases we would like to compute functions on an input with a
particular structure (e.g., a connected graph).  Motivated by this, we
define the classes $\pmud$ and $\pstream$ capturing respectively
mud and streaming algorithms for symmetric functions that are not
necessarily total (they are defined only on inputs that satisfy a
property that is promised).

\begin{definition}
Let $A\subseteq \X^n$.
A symmetric function $f: A \to \X$ is in the class $\pmud$ if there
exists a $\polylog(n)$-communication, $\polylog(n)$-space mud algorithm $m = (\fil, \oplus, \cleanup)$ such that for all $\seq \in A$, and computation trees $\tree$, we have
$\cleanup(m_\tree(\seq)) = f(\seq)$.
\end{definition}

\begin{definition}
Let $A\subseteq \X^n$.
A symmetric function $f: A \to \X$ is in the class $\pstream$ if
there exists a $\polylog(n)$-communication, $\polylog(n)$-space streaming algorithm $s=(\sigma, \cleanup)$ such that for all $\seq \in A$ we have $s^0(\seq) = f(\seq)$.
\end{definition}

\begin{theorem}\label{thm:promisess}
$\pmud \subsetneq \pstream$.
\end{theorem}

To prove Theorem~\ref{thm:promisess},  we introduce a
promise problem, that we call {\sc SymmetricIndex}, and we show that
it is in $\pstream$, but not in $\pmud$.  Intuitively, we want to
define a problem in which the input will consist of two sets of
records.  In the first set, we are given a $n$-bit string
$x_1,\ldots,x_n$, and a query index $p$.  In the second set, we are
given a $n$-bit string $y_1,\ldots,y_n$, and a query index $q$.  We
want to compute either $x_{q}$, or $y_p$, and we are guaranteed that
$x_{q}=y_p$.
Formally, the alphabet of the input is $\X=\{\one,\two\}\times [n] \times \{0,1\} \times [n]$.
An input $S\in \X^{2n}$ is some arbitrary permutation of a sequence
with the form 

\vspace{-.1in}
{\small
\begin{eqnarray*}
S & = &
(\one,1,x_1,p),(\one,2,x_2,p),\ldots,(\one,n,x_n,p),(\two,1,y_1,q),(\two,2,y_2,q),\ldots,(\two,n,y_n,q) .
\end{eqnarray*}}
\vspace{-.1in}

\noindent Additionally, the set $S$ satisfies the promise that 
$x_{q}=y_p$.  
Our task is to compute the function $f(S)=x_{q}$.
In order to prove Theorem~\ref{thm:promisess} we give a deterministic
$\mbox{polylog}(n)$-space streaming algorithm for {\sc
  SymmetricIndex}, and we  show that any deterministic SCM protocol for
the same problem has communication complexity $\Omega(n)$.  Due to
lack of space, the proof appears in the Appendix.

\subsection{Indeterminate Functions}
\label{sec:ind}

In some applications, the function we wish to compute may have more than one ``correct'' answer. 
We define the classes $\imud$ and $\istream$ to capture the computation of ``indeterminate'' functions.

\begin{definition}
A total symmetric function $f: \X^n \to 2^{\X}$ is in the class $\imud$ if there
exists a $\polylog(n)$-communication, $\polylog(n)$-space $\mud$ algorithm $m = (\fil, \oplus, \cleanup)$ such that for all $\seq \in \X^n$, and computation trees $\tree$, we have
$\cleanup(m_\tree(\seq)) \in f(\seq)$.
\end{definition}

\begin{definition}
A total symmetric function $f: \X^n \to 2^{\X}$ is in the class $\istream$ if
there exists a $\polylog(n)$-communication, $\polylog(n)$-space streaming algorithm $s=(\sigma,
\cleanup)$ such that for all $\seq \in X^n$ we have $s^0(\seq) \in  f(\seq)$.
\end{definition}

Consider a promise function $f:A\rightarrow \Sigma$, such that $f\in \pmud$.
We can define a total indeterminate function $f':\Sigma^n\rightarrow 2^{\Sigma}$, such that for each $x\in A$, $f'(x)=f(x)$, and for each $x\notin A$, $f(x)=\Sigma$.
That is, for any input that satisfies the promise of $f$, the two functions are equal, while for all other inputs, any output is acceptable for $f'$.
Clearly, a streaming or mud algorithm for $f'$, is also a streaming or mud  algorithm for $f$ respectively.  Therefore, Theorem~\ref{thm:promisess} implies for following result.

\begin{theorem}\label{thm:approxss}
$\imud \subsetneq \istream$.
\end{theorem}

\section{Concluding Remarks}

Unlike conventional streaming systems that make passes over ordered
data with a single processor, modern log processing systems like
Google's MapReduce~\cite{mapreduce} and Apache's Hadoop~\cite{hadoop}
rely on massive, unordered, distributed (mud) computations to do data
analysis in practice, and get speedups.  Motivated by that, we have
introduced the model of mud algorithms. Our main result is that any symmetric function that
can be computed by a streaming algorithm can be computed by a mud
algorithm as well with comparable space and communication resources,
showing the equivalence of the two classes. At the heart of the proof
is a nondeterministic simulation of a streaming algorithm that guesses
the stream, and an application of Savitch's theorem to be
space-efficient.  This result formalizes some of the intuition that
has been used in designing streaming algorithms in the past
decade. This result has certain natural extensions to approximate and
randomized computations, and we show that other natural extensions to
richer classes of symmetric functions are impossible.

We think the generalization of mud algorithms to reflect the full
power of these modern log processing systems is likely to be a very
exciting area of future research.  In one generalization, a ``multi-key''
mud algorithm computes a function $\X^n \to \X^n$ in a single round,
where each symbol in the output is the result of a ``single-key'' mud
algorithm (as we've defined it in this paper).
%
%
Because generalized mud models already work in practice at
massive scale, algorithmic and complexity-theoretic insights will have
tremendous impact.

There are other technical problems that are open and of interest. In
particular, can one obtain more time-efficient simulation for
Theorem~\ref{thm:main}? Also, D. Sivakumar asked if there are natural
problems for which this simulation provides an interesting
algorithm~\cite{andrew}.

\section*{Acknowledgements}
We thank the anonymous referees for several suggestions to improve a
previous version of this paper, and for suggesting the use of
$\epsilon$-biased generators.  We also thank Sudipto Guha and D. Sivakumar for helpful discussions.

\newpage

\bibliographystyle{plain}
\bibliography{mrfocs}

\begin{thebibliography}{10}

\bibitem{ruhl}
Gagan Aggarwal, Mayur Datar, Sridhar Rajagopalan, and Matthias Ruhl.
\newblock On the streaming model augmented with a sorting primitive.
\newblock In {\em FOCS '04: Proceedings of the 45th Annual IEEE Symposium on
  Foundations of Computer Science (FOCS'04)}, pages 540--549, Washington, DC,
  USA, 2004. IEEE Computer Society.

\bibitem{AMS}
N.~Alon, Y.~Matias, and M.~Szegedy.
\newblock The space complexity of approximating the frequency moments.
\newblock {\em Proceedings of the Symposium on Theory of Computing}, pages
  20--29, 1996.

\bibitem{babai}
L.~Babai, A.~Gal, P.~Kimmel, and S.\ Lokam.
\newblock Simultaneous messages and communication.
\newblock {\em Univ of Chicago, Technical Report}, 1996.

\bibitem{stringeq2}
L.~Babai and P.~G. Kimmel.
\newblock Randomized simultaneous messages: Solution of a problem of yao in
  communication complexity.
\newblock In {\em CCC '97: Proceedings of the 12th Annual IEEE Conference on
  Computational Complexity}, page 239, Washington, DC, USA, 1997. IEEE Computer
  Society.

\bibitem{hadoop}
Andrzej Bialecki, Mike Cafarella, Doug Cutting, and Owen O'Malley.
\newblock Hadoop: a framework for running applications on large clusters built
  of commodity hardware, 2005.
\newblock Wiki at \url{http://lucene.apache.org/hadoop/}.

\bibitem{broder}
A.~Broder, M.~Charikar, A.~Frieze, and M.~Mitzenmacher.
\newblock Min-wise independent permutations.
\newblock {\em J. Comput. Syst. Sci. 60(3)}, pages 630--659, 2000.

\bibitem{datar}
M.~Datar and S.~Muthukrishnan.
\newblock Estimating rarity and similarity over data stream windows.
\newblock {\em ESA}, pages 323--334, 2002.

\bibitem{mapreduce}
Jeffrey Dean and Sanjay Ghemawat.
\newblock Mapreduce: Simplified data processing on large clusters.
\newblock In {\em OSDI'04: Sixth Symposium on Operating System Design and
  Implementation}, 2004.

\bibitem{HRR}
M.~Henzinger, P.~Raghavan, and S.~Rajagopalan.
\newblock Computing on data streams.
\newblock {\em Technical Note 1998-011, Digital Systems Research Center, Palo
  Alto, CA}, 1998.

\bibitem{I}
P.~Indyk.
\newblock Stable distributions, pseudorandom generators, embeddings, and data
  stream computation.
\newblock {\em Journal of ACM}, pages 307--323, 2006.

\bibitem{andrew}
A.~McGregor.
\newblock Open problems in data streams research.
\newblock \url{http://www.cse.iitk.ac.in/users/sganguly/data-stream-probs.pdf}.

\bibitem{muthu}
S.~Muthukrishnan.
\newblock Data streams: Algorithms and applications.
\newblock {\em Foundations and Trends in Theoretical Computer Science}, 2005.

\bibitem{NN}
Joseph Naor and Moni Naor.
\newblock Small-bias probability spaces: Efficient constructions and
  applications.
\newblock {\em SIAM Journal on Computing}, 22(4):838--856, August 1993.

\bibitem{stringeq1}
Ilan Newman and Mario Szegedy.
\newblock Public vs. private coin flips in one round communication games
  (extended abstract).
\newblock In {\em STOC '96: Proceedings of the twenty-eighth annual ACM
  symposium on Theory of computing}, pages 561--570, New York, NY, USA, 1996.
  ACM Press.

\bibitem{saw}
Rob Pike, Sean Dorward, Robert Griesemer, and Sean Quinlan.
\newblock Interpreting the data: Parallel analysis with sawzall.
\newblock {\em Scientific Programming Journal}, 13(4):227--298, 2005.

\bibitem{pavel}
P.~Pudlak, V.~Rodl, and J.~Sgall.
\newblock Boolean circuits, tensor ranks and communication complexity.
\newblock {\em Manuscript}, 1994.

\bibitem{savitch}
Savitch.
\newblock Maze recognizing automata and nondeterministic tape complexity.
\newblock {\em Journal of Computer and System Sciences}, 1973.

\end{thebibliography}

\appendix 

\section{Appendix for Section 3}

\subsection{Proof of Theorem \ref{thm:private_randomness}}
A randomized streaming algorithm for computing $f$ works as follows.
We pick an $\epsilon$-biased family of $n$ binary random variables $X_0,\ldots,X_{n-1}$, for some $\epsilon<1/2$.
Such a family has the property for any $S\subseteq [n]$,
\[
\mathbf{Pr}[\sum_{i\in S}X_i \mbox{ mod } 2 = 1] > 1/4.
\]
Moreover, this family can be constructed using $O(\log{n})$ random bits, such that the value of each $X_i$ can be computed in time $\log^{O(1)}{n}$ \cite{NN}.
We can thus compute in a streaming fashion the bit
$  B  =  b_1 \cdot X_{i_1} + b_2 \cdot X_{i_2} + ... + b_n \cdot X_{i_n}$.
Observe that if $f(S)=1$, then $Pr[B=1] = 0$.
On the other hand, if $f(S)=0$, then let
\[
A = \{t\in \{0,\ldots,n-1\} | \sum_{j:i_j=t} b_j \mbox{ mod } 2 = 1\} .
\]
We have
\[
\mathbf{Pr}[B=1] = \mathbf{Pr}[\sum_{i\in A} X_i \mbox{ mod } 2 = 1] >
1/4 .
\]
Thus, by repeating in parallel the above experiment $O(\log(n))$ times, we obtain a randomized streaming algorithm for {\sc SetParity}, that succeeds with high probability.

It remains to show that there is no SCM protocol for {\sc SetParity} with communication complexity $o(\sqrt{n})$.
We will use a reduction from the string equality problem  \cite{stringeq1, stringeq2}.
Alice gets a string 
$x_1,...,x_n\in \{0,1\}^n$, 
and Bob gets a string 
$y_1,...,y_n\in \{0,1\}^n$.
They independently compute the sets of records
$S_A  =  \{(1,x_1), \ldots, (n,x_n)\}$,
and
$S_B  =  \{(1,y_1), \ldots, (n,y_n)\}$.
It is easy to see that $f(S_A\cup S_B)=1$ iff the answer to the string-equality problem is YES.
Thus, any protocol with private randomness for $f$ has communication complexity $\Omega(\sqrt{n})$.

\subsection{Proof of Theorem \ref{thm:promisess}}
We start by giving a deterministic $\mbox{polylog}(n)$-space streaming
algorithm for {\sc SymmetricIndex} that implies $\mbox{{\sc
SymmetricIndex}}\in \pstream$.  The algorithm is given the elements
of $S$ in an arbitrary order.  If the first record is $(\one,i,x_i,p)$
for some $i$,
the algorithm streams over the remaining records until it gets the
record $(\two,p,y_p,q)$ and outputs $y_p$.  If the first record is
$(\two,j,y_j,q)$ for some $j$, then the algorithm streams over the remaining records
until it gets the record $(\one,q ,x_q,p)$.  In either case we output
$x_q = y_p$.

We next show that $\mbox{{\sc SymmetricIndex}}\notin \pmud$.
It suffices to show that any deterministic SCM protocol for {\sc
SymmetricIndex} requires $\Omega(n)$ bits of communication.  Consider
such a protocol in which Alice and Bob each send $b$ bits to
Carol, and assume for the sake of contradiction that $b<n/40$.
Let $I$ be the set of instances to the {\sc SymmetricIndex} problem,
and simple counting yields that $|I| = n^2 2^{2n-1}$.
For an instance $\phi\in I$, we split it into two pieces $\phi_A$, for
Alice and $\phi_B$, for Bob.  We  assume that these pieces are
$$
\phi_A = (\one,1,x_1^\phi,p^\phi),\ldots,(\one,n,x_n^\phi,p^\phi),
\;\;\ \mbox{and} \;\;\;
\phi_B = (\two,1,y_1^\phi,q^\phi),\ldots,(\two,n,y_n^\phi,q^\phi).
$$
For this partition of the input,
let $I_A$ and $I_B$, be the sets of possible inputs of Alice, and
Bob respectively.
Alice computes a function $h_A:I_A \rightarrow [2^b]$, Bob computes
a function $h_B:I_B\rightarrow [2^b]$, and each sends the result to Carol.
Intuitively, we want to argue that if Alice sends at most $n/40$ bits
to Carol, then for an input that is chosen uniformly at random from
$I$, Carol does not learn the value of $x_i$ for at least some large
fraction of the indices $i$.  We formalize the above intuition with the following Lemma:



\begin{lemma}\label{claim:alice_few_bits}
If we pick $\phi\in I$, and $i\in [n]$ uniformly at random and
independently, then:
\begin{itemize}
\item
With probability at least $4/5$, there exists
$\chi\neq \phi\in I$, such that 
$h_A(\phi_A)=h_A(\chi_A)$,
$p^{\phi}=p^{\chi}$, and
$x_i^\phi \neq x_i^\chi$.
\item
With probability at least $4/5$, there exists
$\psi\neq \phi\in I$, such that 
$h_B(\phi_B)=h_B(\psi_B)$,
$q^{\phi} = q^{\psi}$, and
$y_i^\phi \neq y_i^\psi$.
\end{itemize}
\end{lemma} 

\begin{proof}
Because of the symmetry between the cases for Alice and Bob, it suffices to
prove the assertion for Alice.
For $j\in [2^b]$, $r\in [n]$, let
\[
C_{j,r} = \{\gamma\in I | h_A(\gamma_A)=j \mbox{ and } p^{\gamma} =
r\} .
\]
Let $\alpha_{j,r}$ be the set of indices $t\in [n]$, such that
$x_t^\gamma$ is fixed, for all $\gamma\in C_{j,r}$.
That is,
\[
\alpha_{j,r} = \{t\in [n] | \mbox{ for all } \gamma,\gamma'\in C_{j,r}, x_t^{\gamma} = x_t^{\gamma'} \}.
\]

Observe that if we fix $|\alpha_{j,r}|$ elements $x_i$ in all the
instances in $C_{j,r}$, then any pair $\gamma,\gamma'\in C_{j,r}$ can
differ only in some $x_i$, with $i\notin \alpha_{j,r}$, or in the index
$q$, or in $y_t$, with the constraint that $x_q=y_p$. Thus, for each $j,r\in
[2^b]$,
\begin{eqnarray}
|C_{j,r}|\leq n \cdot 2^{2n-|\alpha_{j,r}|-1}.
\end{eqnarray}
Thus, if $|\alpha_{j,r}|\geq n/20$, then $|C_{j,r}|\leq
n 2^{39n/20-1}$.  
Pick $\phi\in I$, and $i\in [n]$ uniformly at random, and independently,
and let ${\cal E}$ be the event that 
there exists $\chi\neq \phi\in I$, 
such that $h_A(\phi_A)=h_A(\chi_A)$, 
$p^{\phi} = p^{\chi}$, 
and $x_i^\phi\neq x_i^{\chi}$.
Then
\begin{eqnarray*}
\mathbf{Pr}[{\cal E}] & = & 1-\frac{\sum_{j\in [2^b], r\in [n]} |C_{j,r}|\cdot |\alpha_{j,r}|}{n \cdot |I|}\\
 & \geq & 1-\frac{\sum_{j\in [2^b], r\in [n]} n \cdot 2^{n\frac{39}{20}-1} \cdot n + \sum_{j\in [2^b], r\in [n]}|C_{j,r}|\cdot n/20}{n^3 \cdot 2^{2n-1}}\\
 & \geq & 1 - \frac{2^{n/40} \cdot n^3 \cdot 2^{n\frac{39}{20}-1} + n^2 \cdot 2^{2n-1}\cdot n/20}{n^3 \cdot 2^{2n-1}}\\
 & > & 4/5,
\end{eqnarray*}
for sufficiently large $n$.
\end{proof}



Consider an instance $\phi$ chosen uniformly at random from $I$.
Clearly, $p^{\phi}$, and $q^{\phi}$ are distributed uniformly in $[n]$,
$q^{\phi}$, and  $\phi_A$ are independent, and  
$p^{\phi}$, and $\phi_B$ are independent.
Thus, by Lemma~\ref{claim:alice_few_bits} with probability at least
$1-2\left(\frac{1}{5}\right)$ there exist $\chi, \psi\in I$, such that:
\begin{itemize}
\item
$h_A(\phi_A)=h_A(\chi_A)$, 
$p^{\phi}=p^{\chi}$, and
$x_{q^{\phi}}^\phi \neq x_{q^{\phi}}^\chi$.
\item
$h_B(\phi_B)=h_B(\psi_B)$, 
$q^{\phi} = q^{\psi}$, and
${y}_{p^{\phi}}^{\phi} \neq {y}_{p^{\phi}}^{\psi}$.
\end{itemize}
Consider now the instance $\gamma = \chi_A \cup \psi_B$.
That is,
\begin{eqnarray*}
\gamma & = & (\one, 1, x_1^\chi, p^\chi),\ldots,(\one, n, x_n^\chi, p^\chi), (\two, 1, y_1^\psi, q^\psi),\ldots,(\two, n, y_n^\psi, q^\psi)
\end{eqnarray*}
Observe that
\begin{eqnarray*}
x_{q^{\gamma}}^{\gamma} & = & x_{q^{\psi}}^{\chi} \mbox{ (by the definition of $\gamma$)}\\
 & = & x_{q^{\phi}}^{\chi}\\
 & = & 1 - x_{q^{\phi}}^{\phi}\\
 & = & 1 - {y}_{p^{\phi}}^{\phi} \mbox{ (by the promise for $\phi$)}\\
 & = & {y}_{p^{\phi}}^{\psi}\\
 & = & {y}_{p^{\chi}}^{\psi}\\
 & = & {y}_{p^{\gamma}}^{\gamma} \mbox{ (by the definition of $\gamma$)}.
\end{eqnarray*}
Thus, $\gamma$ satisfies the promise of the problem (i.e., $\gamma\in
I$).  Moreover, we have $h_C(h_A(\phi^A), h_B(\phi^B)) =
h_C(h_A(\gamma^A), h_B(\gamma^B))$, while $x_{q^{\phi}}^{\phi}\neq
x_{q^{\gamma}}^{\gamma}$.  It follows that the protocol is not correct.
We have thus shown that $\pmud \subsetneq \pstream$ and proved Theorem~\ref{thm:promisess}.

\end{document}